\documentclass[conference,1opt,final,a4paper]{IEEEtran}
\usepackage{latexsym}
\usepackage{graphicx}
\usepackage{amsfonts,amssymb,amsmath}
\usepackage[binary-units]{siunitx}
\usepackage{xcolor}

\usepackage{tabularx}
\usepackage{multirow}
\usepackage{booktabs}
\usepackage{diagbox}

\DeclareSIUnit{\bps}{bps}

\def\BibTeX{{\rm B\kern-.05em{\sc i\kern-.025em b}\kern-.08em T\kern-.1667em\lower.7ex\hbox{E}\kern-.125emX}}
\begin{document}

\title{An Abstracted Survey on 6G:\\Drivers, Requirements, Efforts, and Enablers}

\author{
	\IEEEauthorblockN{Bin~Han\IEEEauthorrefmark{1}, Wei~Jiang\IEEEauthorrefmark{2},  Mohammad~Asif~Habibi\IEEEauthorrefmark{1}, and Hans~D.~Schotten\IEEEauthorrefmark{1}\IEEEauthorrefmark{2}}
	\IEEEauthorblockA{\IEEEauthorrefmark{1}Technische Universit\"at Kaiserslautern,
		\IEEEauthorrefmark{2}German Research Centre for Artificial Intelligence (DFKI)
	}%
}

\maketitle

\begin{abstract}
	As of today, 5G mobile systems have been already widely rolled out, it is the right time for academia and industry to explore the next generation mobile communication system beyond 5G. To this end, this paper provides an abstracted survey for the 6G mobile system. We shed light on the key driving factors for 6G through predicting the growth trend of mobile traffic and mobile service subscriptions until the year of 2030, envisioning the potential use cases and applications, as well as deriving the potential use scenarios. Then, a number of key performance indicators to support the 6G use cases are identified and their target values are estimated in a quantitatively manner, which is compared with those of 5G clearly in a visualized way.  An investigation of the efforts spent on 6G research in different countries and institutions until now is summarized, and a potential roadmap in terms of the definition, specification, standardization, and spectrum regulation is given. Finally, an introduction to potential key 6G technologies is provided. The principle, technical advantages, challenges, and open research issues for each identified technology are discussed.
\end{abstract}


\section{Introduction}

The year of 2019 opens a new era of the 5G mobile communications. As we are writing the paper, a list of countries such as South Korea, United States, Switzerland, United Kingdom, and Spain have launched commercial 5G services for the general public, while this list grows quickly and is envisioned to become much longer in the near future. As a revolutionary technology, 5G will penetrate into all aspects of society, generating tremendous economic and societal benefits. From the perspective of technology research, however, it is already the time to start considering what the future beyond-5G (B5G) or 6G mobile networks should be, in order to satisfy the demand on communications and networking in 2030. 

Since 2018, several pioneering projects have been launched aiming at the next generation of mobile networks. The International Telecommunication Union Telecommunication Standardization Sector (ITU-T) Focus Group Technologies for Network 2030 (FG NET-2030) was established in July 2018, intending to study the capabilities of networks for the year 2030 and beyond, when it is expected to support novel forward-looking scenarios. The European Commission initiated to sponsor beyond-5G research activities, such as its recent Horizon 2020 call - 5G Long Term Evolution – where a number of pioneer projects will be kicked off at the early beginning of 2020. In Finland, the University of Oulu began ground-breaking 6G research as part of Academy of Finland’s flagship program, 6G-Enabled Wireless Smart Society and Ecosystem (6Genesis). Besides, other counties such as the United States, China, Japan, and South Korea already formally started the research and development of key 6G technologies or at least announced their ambition to support the 6G works. 

During the same time, there has also been a significant literature conducted for 6G, as listed in Tab.~\ref{table_related}. Some of these works focus on the description of use cases or applications, some list a number of potential key technologies but only in rough introduction, while some others are providing detailed technological description to very specific categories of technologies. To the best of our knowledge, yet there is no comprehensive survey that can provide a complete view to link the aforementioned related works into an organic structure. To fill this gap, in this article we attempt to provide a complete and through view of the state-of-the-art advances in the development and research of 6G by providing vision, use cases, use scenarios, requirements, efforts, roadmap, as well as  a introduction to the promising key technologies.

The rest of this article is organized as follows: Sec.~\ref{sec:drivers} discusses the key driving factors for the necessity of developing 6G, including the prediction of the explosive growth of mobile traffic and mobile users by 2030, the envision and use cases and application scenarios. Sec.~\ref{sec:req} analyzes the requirements for the 6G systems by quantitatively depicting a number of key performance indicators. The efforts of research, regulatory, and standardization of the main players in the mobile communication industry are summarized and the potential timelines and roadmap for development and standardization are estimated in Sec.~\ref{sec:enablers}. Sec.~\ref{sec:roadmap} provides a brief survey of a dozen of key technologies that are identified as the key enablers for 6G. Finally, Sec.~\ref{sec:conclusions} concludes this article.

\begin{table*}[!t]
	\renewcommand{\arraystretch}{1.3}
	\caption{A summary of related works in the field of 6G networks}
	\label{table_related}
	\scriptsize
	\centering
	\begin{tabular}{|c|c|m{2cm}|m{12cm}|}
		\hline
		\textbf{Ref.} & \textbf{Time} & \textbf{Topics} & \textbf{Contributions}    \\  \hline \hline
		\cite{Ref_WJ_david20186g}  & Sept. 2018 & Vision & Review the key services and innovations from 1G to 5G, and provide a vision for 6G.\\ \hline
		\cite{Ref_WJ_nawaz2019quantum} & April 2019 & ML & Review the state-of-the-art advances in machine learning and quantum computing, and propose a quantum computing-assisted ML framework for 6G networks. \\ \hline 
		\cite{Ref_WJ_pappaport2019wireless} & June 2019& Terahertz & Describe the technical challenges and potentials for wireless communications and sensing  above $100\mathrm{GHz}$, and presents discoveries, approaches, and recent results that will aid in the development and implementation of the 6G networks.   \\ \hline
		\cite{Ref_WJ_yang2019wireless}& July 2019 & Vision & Outline a number of key technological challenges and the potential solutions associated with 6G. \\ \hline
		\cite{Ref_WJ_Letaief2019roadmap} & Aug. 2019 & AI &  Discuss potential technologies for 6G to enable ubiquitous AI applications and AI-enabled approaches for the design and optimization of 6G.  \\ \hline
		\cite{Ref_WJ_Zong20196g} &Sept. 2019 & Photonics, holography, AI& Proposes two candidate system architectures for 6G and identifies several 6G technologies including photonics-defined radio, holography, and AI.   \\ \hline 
		\cite{Ref_BH_zhang20196g} & Sept. 2019& Survey & A survey aiming to identify requirements, network architecture, key technologies, and new applications.  \\ \hline
		\cite{Ref_WJ_Strinati20196g}& Sept. 2019 & Technological enablers & Five technology enablers for 6G, including pervasive AI at network edge, 3D coverage consisting of terrestrial networks, aerial platforms, and satellite constellation, a new physical layer incorporating sub-Thz and VLC, distributed security mechanisms, and a new architecture. \\ \hline
		\cite{Ref_WJ_huang2019survey} & Dec. 2019 & Green 6G & A survey on new architectural changes associated with 6G networks and potential technologies, such as ubiquitous 3D coverage, pervasive AI, terahertz, visible light communication, and blockchain. \\ \hline
		\cite{Ref_WJ_jiang2019computational} & Dec. 2019 & AI & A special issue provides a comprehensive and highly coherent treatment on all the technology aspects related to ML for wireless communications and networks such as multi-path fading channel, channel coding, and physical-layer design. \\ \hline
		\cite{Ref_WJ_dang2020what} & Jan. 2020 & Vision & Argue that 6G should be human-centric, and therefore security, secrecy, and privacy are key features. To support this vision, a systematic framework, required technologies, and challenges are outlined. \\ \hline
		\cite{Ref_WJ_tang2020future} & Feb. 2020 & Vehicular, ML& A survey on various machine learning technologies that are promising for communication, networking, and security aspects of vehicular networks, and a envision of the ways toward an intelligent 6G vehicular network, including intelligent radio, network intelligentization, and self-learning. \\ \hline
		\cite{Ref_WJ_Giordani2020toward}  & Mar. 2020  & Use cases & Foresee several possible use cases and presents technologies that are considered as the enablers for these 6G use cases.    \\ \hline
		\cite{Ref_WJ_Viswanathan2020communications} & Mar. 2020 & Survey &  New themes including new human-machine interface, ubiquitous computing, multi-sensory data fusion, and precise sensing and actuation, major technology transformations such as new spectrum, new architecture, and new security are presented, and the potential of AI is emphasized. \\ \hline
		\cite{Ref_WJ_chen2020vision} & April 2020 & Survey &  A comprehensive discussion of 6G based on the review of 5G developments, covering visions, requirements, technology trends, and challenges, aiming at clarifying the ways to tackle the challenges of coverage, capacity, data rate, and mobility of 6G communication systems.  \\ \hline
		\cite{Ref_WJ_saad2020vision} & May 2020 & Survey &  A vision on 6G in terms of applications, technological trends, service classes, and requirements, and  an identification on enabling technologies and open research problems.   \\ \hline
		\cite{Ref_WJ_kato2020challenges} & June 2020& ML& Discuss possible challenges and potential research directions of advancing machine learning technologies into the future 6G network in terms of communication, networking, and computing perspective. \\ \hline
		\cite{Ref_WJ_guo2020explainable} & June 2020 & AI & Outline the core concepts of explainable AI for 6G, including public and legal motivations, definitions, trade-off between explainability and performance, and explainable methods, and propose an explainable AI framework for future wireless systems. \\ \hline
	\end{tabular}
\end{table*}

\section{Drivers}\label{sec:drivers}

As of today, the commercial deployment of 5G mobile networks has been rolled out for around one year across the world, and the network scale in some countries is already very large. Following the historical tradition in the mobile industry, i.e., a new generation every ten years, it is the right time to discuss the successor of 5G in the research and standardization community. The key driving forces for the development of a next-generation system are not only from the exponential growth of mobile traffic and connectivity demand, but also from  new disruptive services and applications on the horizon.

\subsection{Explosive Mobile Traffic}

\begin{figure}[!bpht]
\centering
\includegraphics[width=0.4\textwidth]{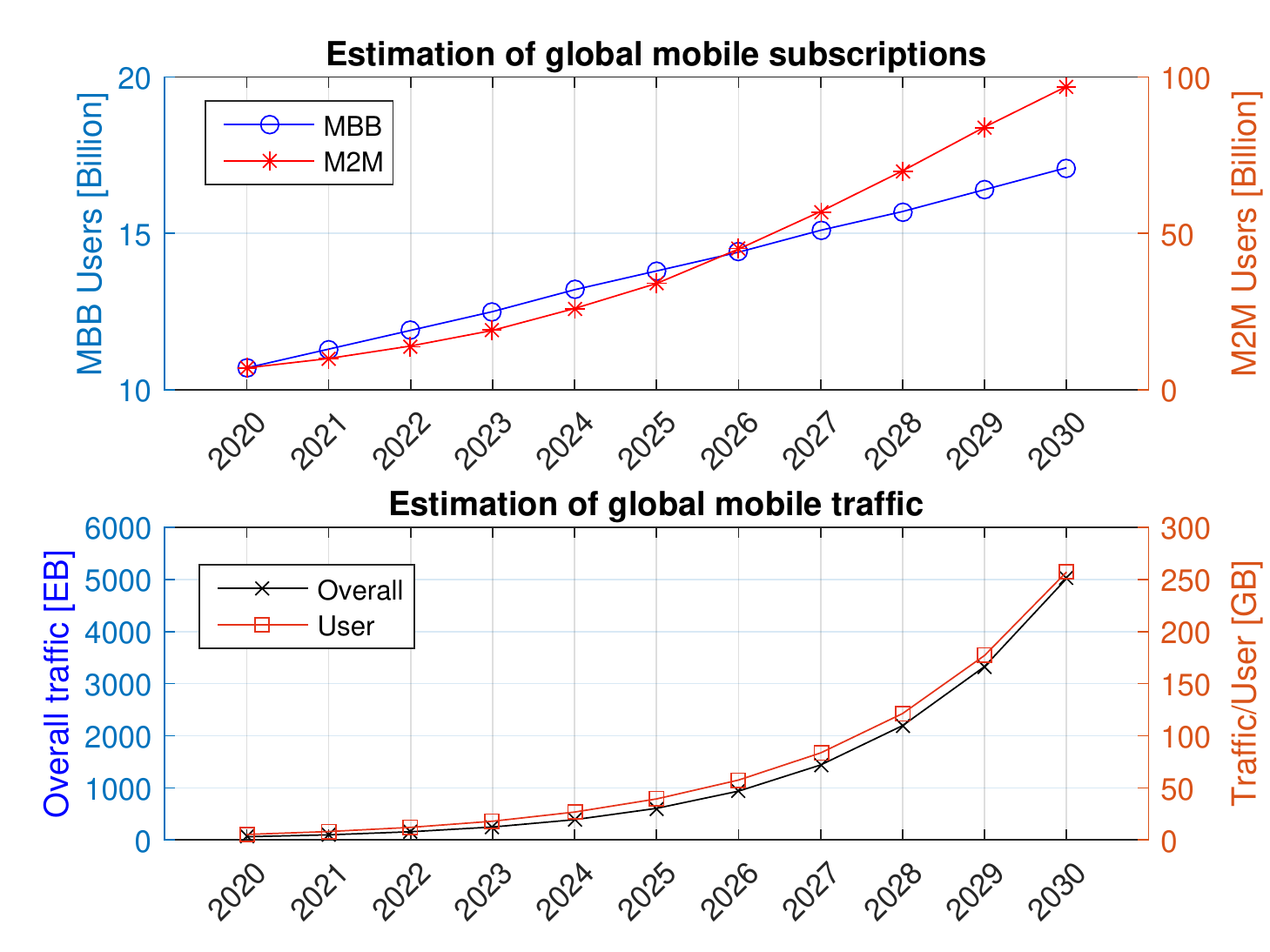}
\caption{The estimated global mobile subscriptions and traffic from 2020 to 2030. Source: ITU-R Report M.2370-0 \cite{Ref_WJ_non2015traffic}. }
\label{Figure_Traffic}
\end{figure}

We are in an unprecedented era where a massive number of smart products, services, and applications emerge and evolve quickly, imposing a huge demand on mobile traffic. It can be foreseen that the 5G system designed before 2020 cannot well satisfy such a demand in 2030 and beyond. Due to the proliferation of rich-video applications, enhanced screen resolution, Machine-to-Machine (M2M) communications, mobile cloud services, etc., the global mobile traffic will continuously increase in an explosive manner, up to \SI{5016}{\exa\byte} per month in the year of 2030 compared with 
\SI{62}{\exa\byte} per month in 2020, according to the estimation of ITU-R made in 2015 \cite{Ref_WJ_non2015traffic}. A report by Ericsson \cite{Ref_WJ_ericsson2020mobile} revealed that the global mobile traffic had reached around \SI{33}{\exa\byte} per month by the end of 2019, which partially proves the correctness of ITU-R's estimation. The number of smartphones and tablets will further increase, while other devices such as wearable electronics will grow in a faster pace, amounting to a total of $17.1$ billion mobile subscriptions in 2030. In addition to the human-centric communications, the M2M terminals will experience a more-rapid growth and will become saturated no earlier than 2030. It is envisioned that the number of M2M subscriptions will reach $97$ billion, as shown in \figurename \ref{Figure_Traffic}, around $14$ times over that of 2020. The traffic from mobile video services already dominated, account for two-thirds of all mobile traffic. However, the usage of video services keeps growing, such as the boom of Tik-Tok recently, and the resolution of video continuously improves. In some developed countries, strong traffic growth before 2025 is driven by rich-video services and  long-term growth wave will continue due to the penetration of Augmented/Virtual Reality (AR/VR) applications. The average data consumption for every mobile user per month, as illustrated in \figurename \ref{Figure_Traffic}, will increase from around \SI{5}{\giga\byte} in 2020 to over \SI{250}{\giga\byte} in 2030.

\subsection{Use cases}
\begin{table*}[!hbtp]
	\centering
	\caption{Some 6G use cases with high potential}
	\label{tab:use_cases}
	\scriptsize
	\begin{tabular}{|l|l|l|}
		\hline
		\textbf{Use Case} & \textbf{Typical Applications} & \textbf{Key Requirements} \\\hline\hline
		Extended reality (ER)&immersive gaming, remote surgery, remote industrial control&high data rate ($\geqslant$\SI{1.6}{\giga\bps}/device), low latency, high reliability\\\hline
		Holographic telepresence&online education, collaborative working, deep-immersive gaming&ultra high data rate (terabits per second)\\\hline
		Multi-sense experience&remote surgery, tactile Internet, remote controlling and reparing&stringently low latency\\\hline
		Tactile Internet for Industry 4.0&industrial automation, smart energy consumption &$\leqslant$\SI{1}{\milli\second} E2E latency\\\hline
		Intell. transport \& logistics& Automated road speed enforcement, real-time parking management &stringently high reliability and low latency\\\hline
		Ubiquitous global roaming&World-wide roaming services for UE, portable devices, industrial apps. &low-cost fully global coverage\\\hline
		Pervasive intelligence&computer vision, SLAM, speech recognition, NLP, motion control& high decision accuracy and transparency, complex data privacy\\\hline
	\end{tabular}
\end{table*}
With the advent and evolution of cutting-edge fields, such as displaying, robotics, edge computing, AI, unmanned aerial vehicle (UAV), and space technology, in combination with the mobile system, many unprecedented use cases can be fostered. Here, we envision several cases with high potential, as summarized in Tab.~\ref{tab:use_cases}.

\begin{figure}[!t]
\centering
\includegraphics[width=0.3\textwidth]{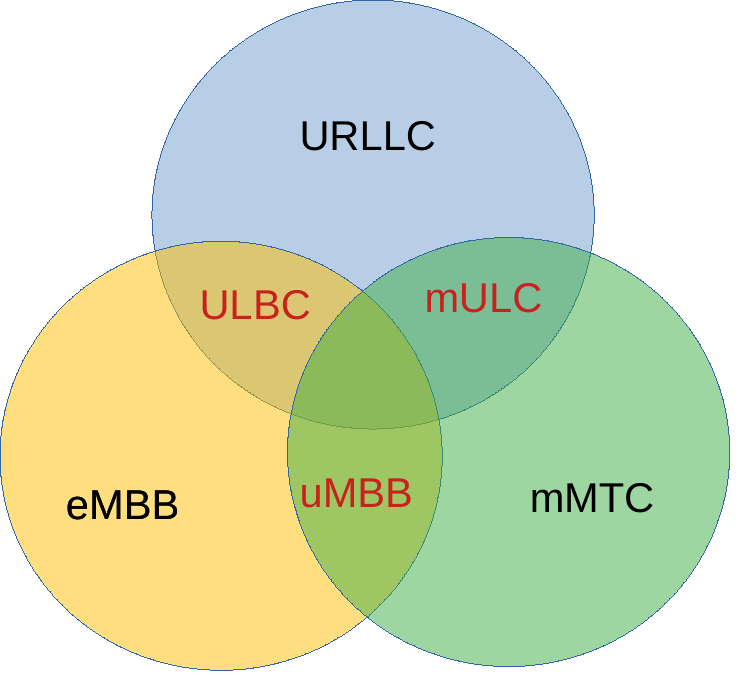}
\caption{The envisioned use scenarios for 6G systems.}
\label{Figure_6GScenarios}
\end{figure}

\subsection{Use scenarios}
In the 5G systems, three usage scenarios have been firstly defined by ITU-R recommendation M.2083 in 2015 \cite{Ref_WJ_non2015imt}. More specifically: the enhanced Mobile Broad-band (eMBB) addresses the human-centric applications for a high-data-rate access to mobile services, multi-media content and data; the Ultra-Reliable Low-Latency Communications (URLLC) focuses on enabling mission-critical connectivity for new applications with stringent requirements on reliability, latency, and availability; while the massive Machine-Type Communications (mMTC) aims at support to dense connectivity with a very large number of connected devices that are low-cost, low-power consumption but typically transmitting a low volume of delay-tolerate data \cite{Ref_MAH_Habibi2019Survey}. 

Being customized for highly specialized applications, all 5G use scenarios achieve extreme performance in some aspect by sacrificing in others, and cannot fully satisfy the requirements of envisioned 6G use cases. For instance, an user wearing a lightweight VR glass playing interactive immersive games requires not only ultra-high data throughput but also very low latency connectivity. Autonomous vehicles on the road or flying drones demand both high data rate and high-reliability low-latency connectivity. Through extending the scope of current 5G use scenarios, as shown in \figurename \ref{Figure_6GScenarios}, we envision four extra scenarios to cover their overlapping areas. 
To accommodate the increasing capacity requirement from commercial passenger planes, helicopters, ships, high-speed trains and the demand of connectivity in remote areas, a ubiquitous coverage of MBB service for the whole planet is expected in 6G, which we named ubiquitous Mobile Broad-Band (uMBB) as an use scenario for 6G. Another key aspect of uMBB is the capacity improvement for hot spots so as to support the proliferation of new services, e.g., VR with a data rate of up to \SI{1}{\giga\bps}/user. Ultra-reliable Low-latency Broadband Communication (ULBC) supports the services with low-latency high-reliability connectivity and high data throughput, such as moving robots and Automatic Guided Vehicle (AGV) in industrial sites. The scenarios of massive Ultra-reliable Low-latency  Communication (mULC) combines the features of mMTC and URLLC, facilitating the deployment of massive sensors and actuators in verticals.

\section{Key Performance Indicator Requirements}\label{sec:req}
\begin{figure}[!t]
\centering
\includegraphics[width=0.4\textwidth]{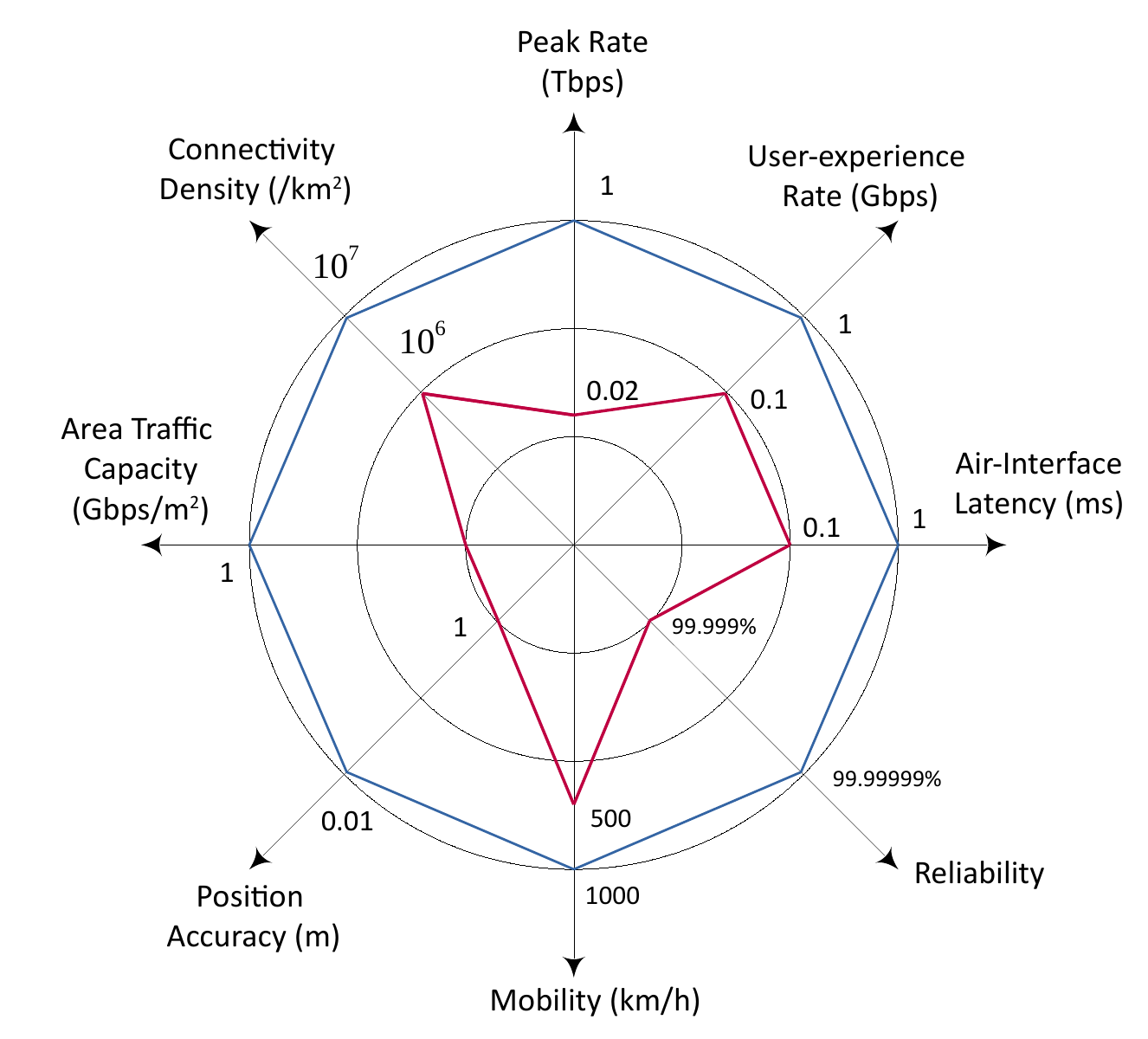}
\caption{The envisioned KPI requirements for 6G in comparison with 5G.}
\label{Figure_5Gflower}
\end{figure}

To satisfy the technical requirements of use scenarios and applications in 2030 and beyond, the 6G system will provide more system capacity and network performance.  Most of the key performance indicators (KPIs) applied for quantitatively or qualitatively evaluating 5G networks are also valid for 6G networks while some new KPIs must be considered for the new features. We briefly summarize our overview to the KPI comparison between 5G and 6G in Fig.~\ref{Figure_5Gflower} and Tab.~\ref{tab:kpi_comp}.

\begin{table*}[!hbtp]
	\centering
	\caption{Comparison between 5G and 6G on some KPI requirements}
	\label{tab:kpi_comp}
	\scriptsize
	\begin{tabular}{|l|c|c|}
		\hline
		\textbf{KPI}&\textbf{5G Requirement}&\textbf{6G Requirement}\\
		\hline\hline
		Peak data rate&\SI{20/10}{\giga\bps} (DL/UL) &\SI{1}{\tera\bps}\\\hline
		\multirow{2}{*}{User-experienced data rate}&\SI{100/50}{\mega\bps}&\multirow{2}{*}{$\geqslant$\SI{1}{\giga\bps}}\\
		& (DL/UL, dense urban)&\\\hline
		\multirow{3}{*}{Latency} &UP: \SI{4/1}{\milli\second} (eMBB/URLLC) & UP: \SIrange{10}{100}{\micro\second}\\
		&CP: \SI{10}{\milli\second} (eMBB/URLLC) & CP: remarkably improved\\
		&E2E: not defined & E2E: considered\\\hline
		Mobility &up to \SI{500}{\kilo\meter/\hour} (high-speed trains)&up to \SI{1000}{\kilo\meter/\hour} (airlines)\\\hline
		Connection density &$10^6$ per \si{\kilo\meter\squared} (with relaxed QoS)&$10^7$ per \si{\kilo\meter\squared}\\\hline
		Network energy efficiency & not defined & 10 -- 100 times better than that of 5G\\\hline
		Peak spectral efficiency &\SI{30/15}{\bps/\hertz} (DL/UL)&\SI{90/45}{\bps/\hertz} (DL/UL)\\\hline
		Area traffic capacity &\SI{10}{\mega\bps/\meter\squared}&\SI{1}{\giga\bps/\meter\squared} (e.g. indoor hot spots)\\\hline
		Reliability &$\geqslant99.999\%$ (URLLC: 32 bytes within \SI{1}{\milli\second}, urban macro)&$\geqslant 99.99999\%$\\\hline
		Signal bandwidth &$\geqslant$\SI{100}{\mega\hertz}&$\geqslant$\SI{1}{\giga\hertz}\\\hline
		Positioning accuracy &$\leqslant$\SI{10}{\meter}&\si{\centi\meter} level\\\hline
		Timeliness &undefined&considered\\\hline
	\end{tabular}
\end{table*}

\section{Roadmap and Efforts}\label{sec:roadmap}
\begin{figure*}[!htbp]
\centering
\includegraphics[width=0.75\textwidth]{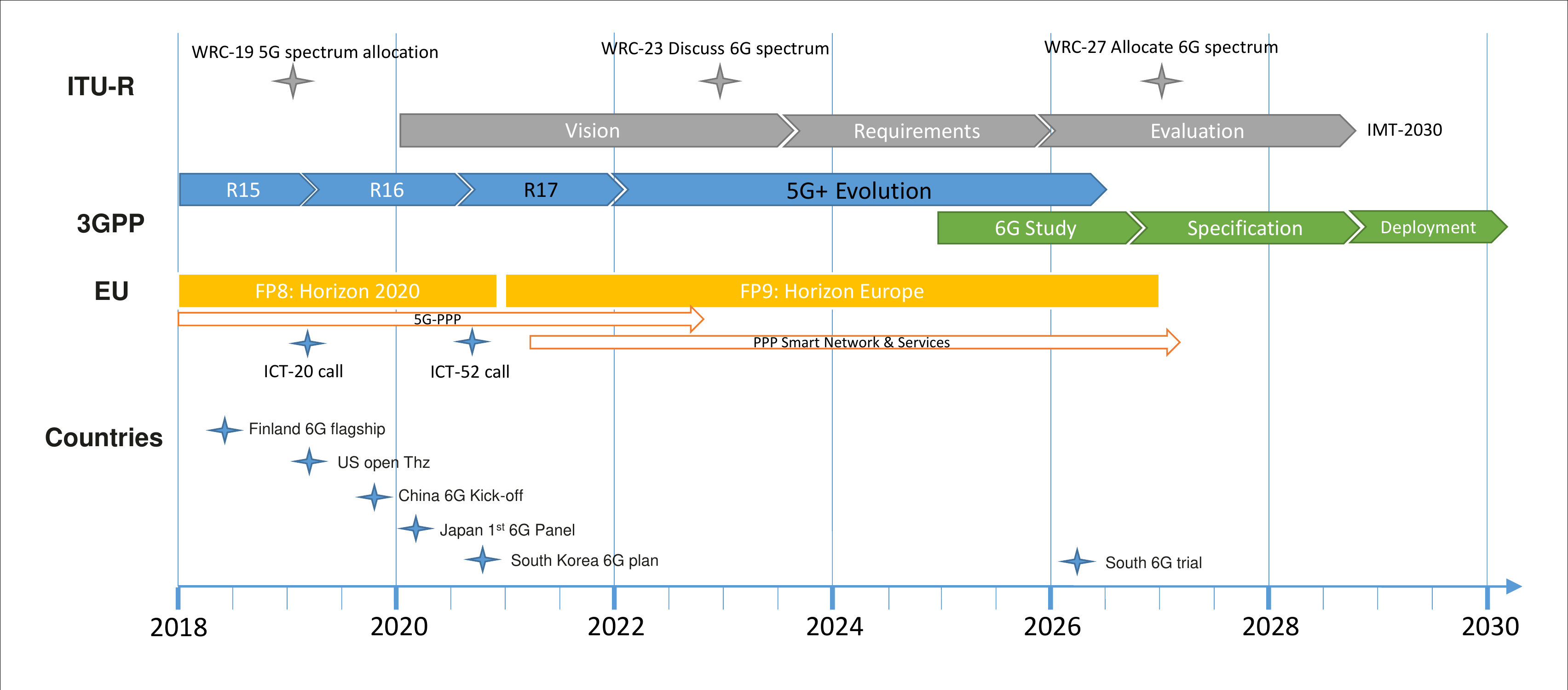}
\caption{The roadmap for main 6G research and developments.} 
\label{Figure_6GTimeline}
\end{figure*}

Although a discussion is ongoing within the wireless community about whether counting should be stop at 5, several pioneering works on the next-generation wireless networks have been initiated, as summarized in Fig.~\ref{Figure_6GTimeline}.

\section{Technological Enablers}\label{sec:enablers}
To pave the road towards the expected extreme performance, so as to realize the envisioned use cases and use scenarios, a diverse set of technologies are expected to play their roles in the future evolution of mobile networks.
We can generally categorize them into the following groups: \textit{New Spectrum} consisting of mmWave, Terahertz communications, Visible Light Communications, Optical Wireless Communications, and the mechanisms of dynamic spectrum management and sharing, \textit{New Networking} covering NFV and SDN, RAN slicing, Open and Smart RAN (O-RAN), and Security, \textit{New Air Interface} that including massive MIMO, Intelligent Reflecting Surfaces, Coordinated Multi-Point, Cell-Free massive MIMO, and new waveform modulation techniques, \textit{New Architecture} providing 3D coverage by means of integrating large-scale satellite constellation, High-Altitude Platform, and Unmanned Aerial Vehicle with traditional terrestrial networks, and  \textit{New Paradigm} empowered by Artificial Intelligence, blockchain, Digital Twin, and Communication-Computing-Control (CoCoCo) convergence. A brief but systematic summary to them is given in Tab.~\ref{tab:key_enablers}.

\begin{table*}[!hbtp]
	\centering
	\caption{Categorized key enablers with advantages and challenges}
	\label{tab:key_enablers}
	\scriptsize
	\begin{tabular}{|l|m{1cm}l|l|l|}
		\hline
		\textbf{Category}&\multicolumn{2}{l|}{\textbf{Enabler}}&\textbf{Advantages}&\textbf{Challenges}\\
		\hline\hline
		\multirow{5}{*}{Spectrum}&\multicolumn{2}{l|}{mmWave}&high bandwidth, narrow beams, high integration&\multirow{2}{*}{severe attenuation \& blockage, low range}\\\cline{2-3}
		&\multicolumn{2}{l|}{THz}&level \cite{Ref_BH_zhang20196g,Ref_BH_wang2018millimeter,Ref_BH_chen2019survey}&\\\cline{2-5}
		&\multicolumn{2}{l|}{OWC}&almost unlimited bandwidth, license-free, low cost, &frail MIMO gain, HW implementation, noise, loss,\\
		&\multicolumn{2}{l|}{(incl. VLC)}&security, health-friendly& nonlinearity, dispersion, pointing errors \cite{Ref_BH_pathak2015visible,Ref_WJ_kahn1997wireless}\\\cline{2-5}
		&\multicolumn{2}{l|}{DSM}&coexistence of licensed/unlicensed users&all-spec. sensing, data processing\&mngmt \cite{Ref_BH_akyildiz20206g,Ref_BH_kliks2020beyond}\\\hline
		\multirow{6}{*}{Networking}&&&&service heterogeneity, SDN controller placement, auto\\
		&\multicolumn{2}{l|}{NFV \& SDN}&high flexibility, low operational cost \cite{Ref_WJ_jiang2017autonomic} &network management \cite{Ref_WJ_jiang2017experimental} and orchestration \cite{Ref_WJ_jiang2017intelligent}, E2E\\
		&&&&QoS control\cite{Ref_MAH_Ksentini2016SDN, Ref_MAH_Alevizaki2020SDN,Ref_MAH_Khan2017SDN,Ref_MAH_Tomovic2019SDN}\\\cline{2-5}
		&\multicolumn{2}{l|}{RAN Slicing}&flexibility, resource efficiency, security&arch. framework supporting multi-use-case verticals \cite{Ref_MAH_Habibi2020SlicingMultiple} \\\cline{2-5}
		&\multicolumn{2}{l|}{O-RAN}&efficiency, intelligence, flexibility, dynamic&lack of tech. convergence \& standardization efforts\\\cline{2-5}
		&\multicolumn{2}{l|}{Security}&service-based E2E security&AI \& ML deployment\\\hline
		\multirow{6}{*}{Air interface}&&&high capacity, statistical multiplexing gain& extremely-large aperture, channel prediction \cite{Ref_WJ_jiang2018multi, Ref_WJ_jiang2019recurrent},\\
		&\multicolumn{2}{l|}{Massive MIMO}& high spec. efficiency, low expenditure,&\cite{Ref_WJ_jiang2020recurrent}, intell. envir. aware adaptation, holographic mMIMO,\\
		&&&low energy consumption&6D positioning, large-scale MIMO radar \cite{Ref_MAH_Huang2020MIMO,Ref_MAH_Bjornson2019MIMO}\\\cline{2-5}
		&\multicolumn{2}{l|}{IRS, aka. RIS/SRE}&high MIMO gain, low imp. cost, low power~\cite{Ref_BH_mossallamy2020reconfigurable,Ref_BH_renzo2020smart}&rely on 3\textsuperscript{rd} party assessments, business framework\\\cline{2-5}
		&\multicolumn{2}{l|}{CoMP}&BS-level spatial diversity, ``cell-free'' potential&clustering, sync., channel est., backhaul\cite{Ref_BH_song2018kpi}\\\cline{2-5}
		&New&\multicolumn{1}{|l|}{Intell. OFDMA}&online MIMO precoding \& resource mapping&waveform design \cite{Ref_WJ_jiang2015ofdm}, out-of-band radiation \cite{Ref_WJ_jiang2012suppressing} \\\cline{3-5}
		&modulation&\multicolumn{1}{|l|}{NOMA}&high power \& spec. efficiencies&specific D2D interface \cite{Ref_WJ_jiang2017device} for cooperative decoding\\\hline
		\multirow{5}{*}{Architecture}&\multicolumn{2}{l|}{Large-scale LEO satellite}&ubiquitous coverage, resistance to natural disasters,&\multirow{2}{*}{integration with terrestrial networks, launching cost}\\
		&constellation&&lower channel loss and cost than GEO  \cite{Ref_WJ_qu2017leo,Ref_WJ_hu2001satellite} &\\\cline{2-5}
		&\multirow{2}{*}{HAP}&& large coverage, unobstructed, flexible deployment;&channel modelling \cite{Ref_WJ_jiang2014opportunistic}, deployment, path planning, opera-\\
		&&&lower cost, easier access to infrastructure and better&tional altitude, interference, energy limit, reliability \cite{Ref_WJ_jiang2014achieving} \\\cline{2-5}
		&\multicolumn{2}{l|}{UAV}&channel than satellite, new use scenarios \cite{Ref_WJ_Mohammed2011role, Ref_WJ_jiang2014opportunistic}&security, real-time demand \cite{Ref_BH_li2019uav}\\\hline
		\multirow{12}{*}{Paradigm} & \multirow{4}{*}{AI}& \multicolumn{1}{|l|}{Deep learning}&automatic featuring \cite{Ref_WJ_jiang2020deep}, prediction \cite{Ref_WJ_jiang2020deeplearning} & computational complexity \cite{Ref_WJ_jiang2019neural} \\\cline{3-5}
		&& \multicolumn{1}{|l|}{Federated learning}&protection to data privacy&communication overheads, heterogeneity \cite{Ref_BH_li2020feaderated}\\\cline{3-5}
		&&\multicolumn{1}{|l|}{Transfer learning}& quick model adaptation and optimization at local&unification, gain measuring, dataset dissimilarity\\\cline{3-5}
		&&\multicolumn{1}{|l|}{AI as a service}&low latency AI service for end-user at terminals&new distributed AI techniques, new APIs\\\cline{2-5}
		&\multirow{2}{*}{Block-chain}&&immutability, decentralization, transparency, security&majority vulnerability, double-spending, transaction\\
		&&&and privacy \cite{Ref_WJ_dai2019blockchain}&privacy leakage, scalability, quantum computing \cite{Ref_BH_nguyen2020privacy}\\\cline{2-5}
		&\multirow{2}{*}{Digital twin}&&improved quality of products, services, processes,&scalability, self-management, lack of models and\\
		&&&devices, etc. in Industry 4.0 and IoT&methodologies, security and privacy \cite{Ref_MAH_Minerva2020DitialTwin,Ref_MAH_Pires2019DigitalTwin,Ref_MAH_Barricelli2019DigitalTwin}\\\cline{2-5}
		&\multirow{2}{*}{Edge intelligence}&&resolves MEC issue caused by service requirement&customized AI algorithms, resource management\\
		&&&diversity among a massive number of users&and task scheduling \cite{Ref_BH_hu2020edge}\\\cline{2-5}
		&&&resolves timeliness and resilience problems due to&\\
		&\multicolumn{2}{l|}{CoCoCo convergence}& the coupling between communication, computation,&in-loop co-design methodologies \& frameworks\\
		&&&and control systems \cite{Ref_BH_han2020robustness, Ref_BH_jiang2020ai}&\\\hline
	\end{tabular}
\end{table*}

\section{Conclusions}\label{sec:conclusions}
In this paper, we provided an abstracted outlook at the drivers, requirements, efforts, and enablers for the next-generation mobile system beyond 5G. The prediction of the trends, the envision of the future societal and technological evaluation, and the identification of key research directions might be rough, partial and even somehow inaccurate with the limitation of the knowledge of the authors and the information we can collect up to now. 

\bibliographystyle{IEEEtran}
\bibliography{Ref_6GReview}

\begin{thebibliography}{10}
\providecommand{\url}[1]{#1}
\csname url@samestyle\endcsname
\providecommand{\newblock}{\relax}
\providecommand{\bibinfo}[2]{#2}
\providecommand{\BIBentrySTDinterwordspacing}{\spaceskip=0pt\relax}
\providecommand{\BIBentryALTinterwordstretchfactor}{4}
\providecommand{\BIBentryALTinterwordspacing}{\spaceskip=\fontdimen2\font plus
\BIBentryALTinterwordstretchfactor\fontdimen3\font minus
  \fontdimen4\font\relax}
\providecommand{\BIBforeignlanguage}[2]{{%
\expandafter\ifx\csname l@#1\endcsname\relax
\typeout{** WARNING: IEEEtran.bst: No hyphenation pattern has been}%
\typeout{** loaded for the language `#1'. Using the pattern for}%
\typeout{** the default language instead.}%
\else
\language=\csname l@#1\endcsname
\fi
#2}}
\providecommand{\BIBdecl}{\relax}
\BIBdecl

\bibitem{Ref_WJ_david20186g}
K.~David and H.~Berndt, ``{6G} vision and requirements: Is there any need for
  beyond {5G}?'' \emph{IEEE Veh. Technol. Mag.}, vol.~13, no.~3, pp. 72--80,
  Sep. 2018.

\bibitem{Ref_WJ_nawaz2019quantum}
S.~J. Nawaz \emph{et~al.}, ``Quantum machine learning for {6G} communication
  networks: State-of-the-art and vision for the future,'' \emph{IEEE Access},
  vol.~7, pp. 46\,317--46\,350, Apr. 2019.

\bibitem{Ref_WJ_pappaport2019wireless}
T.~S. Rappaport \emph{et~al.}, ``Wireless communications and applications above
  100 {GHz}: Opportunities and challenges for {6G} and beyond,'' \emph{IEEE
  Access}, vol.~7, pp. 78\,729--78\,757, Jun. 2019.

\bibitem{Ref_WJ_yang2019wireless}
P.~Yang \emph{et~al.}, ``{6G} wireless communications: Vision and potential
  techniques,'' \emph{IEEE Network}, vol.~33, no.~4, pp. 70--75, Jul. 2019.

\bibitem{Ref_WJ_Letaief2019roadmap}
K.~B. Letaief \emph{et~al.}, ``The roadmap to {6G}: {AI} empowered wireless
  networks,'' \emph{IEEE Commun. Mag.}, vol.~57, no.~8, pp. 84--90, Aug. 2019.

\bibitem{Ref_WJ_Zong20196g}
B.~Zong \emph{et~al.}, ``{6G} technologies: Key drivers, core requirements,
  system architectures, and enabling technologies,'' \emph{IEEE Veh. Technol.
  Mag.}, vol.~14, no.~3, pp. 18--27, Sep. 2019.

\bibitem{Ref_BH_zhang20196g}
Z.~{Zhang}, Y.~{Xiao}, Z.~{Ma}, M.~{Xiao}, Z.~{Ding}, X.~{Lei}, G.~K.
  {Karagiannidis}, and P.~{Fan}, ``{6G} wireless networks: Vision,
  requirements, architecture, and key technologies,'' \emph{IEEE Vehicular
  Technology Magazine}, vol.~14, no.~3, pp. 28--41, 2019.

\bibitem{Ref_WJ_Strinati20196g}
E.~C. Strinati \emph{et~al.}, ``{6G}: The next frontier: From holographic
  messaging to artificial intelligence using subterahertz and visible light
  communication,'' \emph{IEEE Veh. Technol. Mag.}, vol.~14, no.~3, pp. 42--50,
  Sep. 2019.

\bibitem{Ref_WJ_huang2019survey}
T.~Huang \emph{et~al.}, ``A survey on green {6G} network: Architecture and
  technologies,'' \emph{IEEE Access}, vol.~7, pp. 175\,758--175\,768, Dec.
  2019.

\bibitem{Ref_WJ_jiang2019computational}
W.~Jiang and F.-L. Luo, ``Computational radio intelligence: One key for {6G}
  wireless,'' \emph{ZTE Commun.}, vol.~17, no.~4, pp. 1--3, Dec. 2019.

\bibitem{Ref_WJ_dang2020what}
S.~Dang \emph{et~al.}, ``What should {6G} be?'' \emph{Nat. Electron.}, vol.~3,
  p. 20–29, Jan. 2020.

\bibitem{Ref_WJ_tang2020future}
F.~Tang \emph{et~al.}, ``Future intelligent and secure vehicular network toward
  {6G}: Machine-learning approaches,'' \emph{Proc. IEEE}, vol. 108, no.~2, pp.
  292--307, Feb. 2020.

\bibitem{Ref_WJ_Giordani2020toward}
M.~Giordani \emph{et~al.}, ``Toward {6G} networks: Use cases and
  technologies,'' \emph{IEEE Commun. Mag.}, vol.~58, no.~3, pp. 55--61, Mar.
  2020.

\bibitem{Ref_WJ_Viswanathan2020communications}
H.~Viswanathan and P.~E. Mogensen, ``Communications in the {6G} era,''
  \emph{IEEE Access}, vol.~8, pp. 57\,063--57\,074, Mar. 2020.

\bibitem{Ref_WJ_chen2020vision}
S.~Chen \emph{et~al.}, ``Vision, requirements, and technology trend of {6G}:
  How to tackle the challenges of system coverage, capacity, user data-rate and
  movement speed,'' \emph{IEEE Wireless Commun. Mag.}, vol.~27, no.~2, pp.
  218--228, Apr. 2020.

\bibitem{Ref_WJ_saad2020vision}
W.~Saad \emph{et~al.}, ``A vision of {6G} wireless systems: Applications,
  trends, technologies, and open research problems,'' \emph{IEEE Network},
  vol.~34, no.~3, pp. 134--142, May 2020.

\bibitem{Ref_WJ_kato2020challenges}
N.~Kato \emph{et~al.}, ``Ten challenges in advancing machine learning
  technologies toward {6G},'' \emph{IEEE Wireless Commun. Mag.}, vol.~27,
  no.~3, pp. 96--103, Jun. 2020.

\bibitem{Ref_WJ_guo2020explainable}
W.~Guo, ``Explainable artificial intelligence for {6G}: Improving trust between
  human and machine,'' \emph{IEEE Commun. Mag.}, vol.~58, no.~6, pp. 39--45,
  Jun. 2020.

\bibitem{Ref_WJ_non2015traffic}
\emph{{IMT} traffic estimates for the years 2020 to 2030}, ITU-R Std. M.2370-0,
  Jul. 2015.

\bibitem{Ref_WJ_ericsson2020mobile}
``Mobile data traffic outlook,'' Report, Ericsson, Jun. 2020.

\bibitem{Ref_WJ_non2015imt}
\emph{{IMT} vision-Framework and overall objectives of the future development
  of {IMT} for 2020 and beyond}, ITU-R Std. M.2083-0, Sep. 2015.

\bibitem{Ref_MAH_Habibi2019Survey}
M.~A. {Habibi}, M.~{Nasimi}, B.~{Han}, and H.~D. {Schotten}, ``A comprehensive
  survey of ran architectures toward 5g mobile communication system,''
  \emph{IEEE Access}, vol.~7, pp. 70\,371--70\,421, 2019.

\bibitem{Ref_BH_wang2018millimeter}
X.~{Wang}, L.~{Kong}, F.~{Kong}, F.~{Qiu}, M.~{Xia}, S.~{Arnon}, and G.~{Chen},
  ``Millimeter wave communication: A comprehensive survey,'' \emph{IEEE
  Communications Surveys Tutorials}, vol.~20, no.~3, pp. 1616--1653, 2018.

\bibitem{Ref_BH_chen2019survey}
Z.~{Chen}, X.~{Ma}, B.~{Zhang}, Y.~{Zhang}, Z.~{Niu}, N.~{Kuang}, W.~{Chen},
  L.~{Li}, and S.~{Li}, ``A survey on terahertz communications,'' \emph{China
  Communications}, vol.~16, no.~2, pp. 1--35, 2019.

\bibitem{Ref_BH_pathak2015visible}
P.~H. {Pathak}, X.~{Feng}, P.~{Hu}, and P.~{Mohapatra}, ``Visible light
  communication, networking, and sensing: A survey, potential and challenges,''
  \emph{IEEE Communications Surveys Tutorials}, vol.~17, no.~4, pp. 2047--2077,
  2015.

\bibitem{Ref_WJ_kahn1997wireless}
J.~M. Kahn and J.~R. Barry, ``Wireless infrared communications,''
  \emph{Proceedings of the IEEE}, vol.~85, no.~2, pp. 265--298, Feb. 1997.

\bibitem{Ref_BH_akyildiz20206g}
I.~F. {Akyildiz}, A.~{Kak}, and S.~{Nie}, ``6g and beyond: The future of
  wireless communications systems,'' \emph{IEEE Access}, vol.~8, pp.
  133\,995--134\,030, 2020.

\bibitem{Ref_BH_kliks2020beyond}
A.~{Kliks}, L.~{Kulacz}, P.~{Kryszkiewicz}, H.~{Bogucka}, M.~{Dryjanski},
  M.~{Isaksson}, G.~P. {Koudouridis}, and P.~{Tengkvist}, ``Beyond {5G}: Big
  data processing for better spectrum utilization,'' \emph{IEEE Vehicular
  Technology Magazine}, vol.~15, no.~3, pp. 40--50, 2020.

\bibitem{Ref_WJ_jiang2017autonomic}
W.~Jiang, M.~Strufe, and H.~Schotten, ``Autonomic network management for
  software-define and virtualized {5G} systems,'' in \emph{Proc. European
  Wireless}, Dresden, Germany, May 2017.

\bibitem{Ref_WJ_jiang2017experimental}
W.~Jiang, M.~Strufe, and H.~D. Schotten, ``Experimental results for artificial
  intelligence-based self-organized {5G} networks,'' in \emph{Proc. {IEEE} Int.
  Symp. on Personal, Indoor and Mobile Radio Commun. (PIMRC)}, Montreal, QC,
  Canada, Oct. 2017.

\bibitem{Ref_WJ_jiang2017intelligent}
W.~Jiang, M.~Strufe, and H.~Schotten, ``Intelligent network management for {5G}
  systems: The {SELFNET} approach,'' in \emph{Proc. {IEEE} Eur. Conf. on Net.
  and Commun. (EUCNC)}, Oulu, Finland, Jun. 2017, pp. 109--113.

\bibitem{Ref_MAH_Ksentini2016SDN}
A.~{Ksentini}, M.~{Bagaa}, and T.~{Taleb}, ``On using {SDN} in {5G}: The
  controller placement problem,'' in \emph{2016 IEEE Global Communications
  Conference (GLOBECOM)}, 2016, pp. 1--6.

\bibitem{Ref_MAH_Alevizaki2020SDN}
V.-M. Alevizaki, M.~Anastasopoulos, A.~Tzanakaki, and D.~Simeonidou, ``Joint
  fronthaul optimization and {SDN} controller placement in dynamic {5G}
  networks,'' in \emph{Optical Network Design and Modeling}, A.~Tzanakaki,
  M.~Varvarigos, R.~Mu{\~{n}}oz, R.~Nejabati, N.~Yoshikane, M.~Anastasopoulos,
  and J.~Marquez-Barja, Eds.\hskip 1em plus 0.5em minus 0.4em\relax Cham:
  Springer International Publishing, 2020, pp. 181--192.

\bibitem{Ref_MAH_Khan2017SDN}
S.~{Khan}, A.~{Gani}, A.~W. {Abdul Wahab}, M.~{Guizani}, and M.~K. {Khan},
  ``Topology discovery in software defined networks: Threats, taxonomy, and
  state-of-the-art,'' \emph{IEEE Communications Surveys Tutorials}, vol.~19,
  no.~1, pp. 303--324, 2017.

\bibitem{Ref_MAH_Tomovic2019SDN}
S.~{Tomovic} and I.~{Radusinovic}, ``Toward a scalable, robust, and {QoS}-aware
  virtual-link provisioning in {SDN}-based {ISP} networks,'' \emph{IEEE
  Transactions on Network and Service Management}, vol.~16, no.~3, pp.
  1032--1045, 2019.

\bibitem{Ref_MAH_Habibi2020SlicingMultiple}
M.~A. {Habibi}, B.~{Han}, F.~Z. {Yousaf}, and H.~D. {Schotten}, ``How should
  network slice instances be provided to multiple use cases of a single
  vertical industry?'' \emph{IEEE Communications Standards Magazine}, vol.~4,
  no.~3, pp. 53--61, 2020.

\bibitem{Ref_WJ_jiang2018multi}
W.~Jiang and H.~D. Schotten, ``Multi-antenna fading channel prediction
  empowered by artificial intelligence,'' in \emph{Proc. {IEEE} Veh. Tech.
  Conf. (VTC)}, Chicago, USA, Aug. 2018.

\bibitem{Ref_WJ_jiang2019recurrent}
W.~Jiang and H.~Schotten, ``Recurrent neural network-based frequency-domain
  channel prediction for wideband communications,'' in \emph{Proc. {IEEE} Veh.
  Tech. Conf. (VTC)}, Kuala Lumpur, Malaysia, Apr. 2019.

\bibitem{Ref_WJ_jiang2020recurrent}
W.~Jiang and H.~D. Schotten, ``Recurrent neural networks with long short-term
  memory for fading channel prediction,'' in \emph{Proc. {IEEE} Veh. Tech.
  Conf. (VTC)}, Antwerp, Belgium, May 2020.

\bibitem{Ref_MAH_Huang2020MIMO}
C.~{Huang}, S.~{Hu}, G.~C. {Alexandropoulos}, A.~{Zappone}, C.~{Yuen},
  R.~{Zhang}, M.~{Di Renzo}, and M.~{Debbah}, ``Holographic mimo surfaces for
  6g wireless networks: Opportunities, challenges, and trends,'' \emph{IEEE
  Wireless Communications}, pp. 1--8, 2020.

\bibitem{Ref_MAH_Bjornson2019MIMO}
E.~{Björnson}, L.~{Sanguinetti}, H.~{Wymeersch}, J.~{Hoydis}, and T.~L.
  {Marzetta}, ``Massive mimo is a reality—what is next?: Five promising
  research directions for antenna arrays,'' \emph{ScienceDirect: Digital Signal
  Processing}, vol.~94, pp. 3--20, 2019.

\bibitem{Ref_BH_mossallamy2020reconfigurable}
M.~A. {El Mossallamy}, H.~{Zhang}, L.~{Song}, K.~G. {Seddik}, Z.~{Han}, and
  G.~Y. {Li}, ``Reconfigurable intelligent surfaces for wireless
  communications: Principles, challenges, and opportunities,'' \emph{IEEE
  Transactions on Cognitive Communications and Networking}, pp. 1--1, 2020.

\bibitem{Ref_BH_renzo2020smart}
M.~D. {Renzo}, A.~{Zappone}, M.~{Debbah}, M.~{Alouini}, C.~{Yuen}, J.~D.
  {Rosny}, and S.~{Tretyakov}, ``Smart radio environments empowered by
  reconfigurable intelligent surfaces: How it works, state of research, and
  road ahead,'' \emph{IEEE Journal on Selected Areas in Communications}, pp.
  1--1, 2020.

\bibitem{Ref_BH_song2018kpi}
G.~{Song}, W.~{Wang}, D.~{Chen}, and T.~{Jiang}, ``Kpi/kqi-driven coordinated
  multipoint in 5g: Measurements, field trials, and technical solutions,''
  \emph{IEEE Wireless Communications}, vol.~25, no.~5, pp. 23--29, 2018.

\bibitem{Ref_WJ_jiang2015ofdm}
W.~Jiang and T.~Kaiser, ``From {OFDM} to {FBMC}: Principles and
  {Comparisons},'' in \emph{Signal Processing for 5G: Algorithms and
  Implementations}, F.~L. Luo and C.~Zhang, Eds.\hskip 1em plus 0.5em minus
  0.4em\relax United Kindom: John Wiley\&Sons and IEEE Press, 2016, ch.~3.

\bibitem{Ref_WJ_jiang2012suppressing}
W.~Jiang and M.~Schellmann, ``Suppressing the out-of-band power radiation in
  multi-carrier systems: A comparative study,'' in \emph{Proc. {IEEE} Glob.
  Commun. Conf. {(Globecom)}}, Anaheim, USA, Dec. 2012, pp. 1477--1482.

\bibitem{Ref_WJ_jiang2017device}
W.~Jiang, ``Device-to-device based cooperative relaying for {5G} network: A
  comparative review,'' \emph{ZTE Commun.}, vol.~15, no.~S1, pp. 60--66, Jun.
  2017.

\bibitem{Ref_WJ_qu2017leo}
Z.~Qu \emph{et~al.}, ``{LEO} satellite constellation for internet of things,''
  \emph{IEEE Access}, vol.~5, pp. 18\,391--18\,401, Sep. 2017.

\bibitem{Ref_WJ_hu2001satellite}
Y.~Hu and V.~O.~K. Li, ``Satellite-based internet: A tutorial,'' vol.~39,
  no.~3, pp. 154--162, Mar. 2001.

\bibitem{Ref_WJ_jiang2014opportunistic}
W.~Jiang \emph{et~al.}, ``Opportunistic relaying over aerial-to-terrestrial and
  device-to-device radio channels,'' in \emph{Proc. {IEEE} Int. Conf. on
  Commun. {(ICC)}}, Sydney, Australia, Jul. 2014, pp. 206--211.

\bibitem{Ref_WJ_jiang2014achieving}
------, ``Achieving high reliability in aerial-terrestrial networks:
  Opportunistic space-time coding,'' in \emph{Proc. {IEEE} Eur. Conf. on Net.
  and Commun. (EUCNC)}, Bologne, Italy, Jun. 2014.

\bibitem{Ref_WJ_Mohammed2011role}
A.~Mohammed \emph{et~al.}, ``The role of high-altitude platforms ({HAPs}) in
  the global wireless connectivity,'' \emph{Proceedings of IEEE}, vol.~99,
  no.~11, pp. 1939--1953, Nov. 2011.

\bibitem{Ref_BH_li2019uav}
B.~{Li}, Z.~{Fei}, and Y.~{Zhang}, ``Uav communications for 5g and beyond:
  Recent advances and future trends,'' \emph{IEEE Internet of Things Journal},
  vol.~6, no.~2, pp. 2241--2263, 2019.

\bibitem{Ref_WJ_jiang2020deep}
W.~Jiang and H.~Schotten, ``Deep learning for fading channel prediction,''
  \emph{IEEE Open J. of the Commun. Society}, vol.~1, pp. 320--332, Mar. 2020.

\bibitem{Ref_WJ_jiang2020deeplearning}
W.~Jiang and H.~D. Schotten, ``A deep learning method to predict fading channel
  in multi-antenna systems,'' in \emph{Proc. {IEEE} Veh. Tech. Conf. (VTC)},
  Antwerp, Belgium, May 2020.

\bibitem{Ref_WJ_jiang2019neural}
------, ``Neural network-based fading channel prediction: A comprehensive
  overview,'' \emph{IEEE Access}, vol.~7, pp. 118\,112--118\,124, Aug. 2019.

\bibitem{Ref_BH_li2020feaderated}
T.~{Li}, A.~K. {Sahu}, A.~{Talwalkar}, and V.~{Smith}, ``Federated learning:
  Challenges, methods, and future directions,'' \emph{IEEE Signal Processing
  Magazine}, vol.~37, no.~3, pp. 50--60, 2020.

\bibitem{Ref_WJ_dai2019blockchain}
H.-N. Dai \emph{et~al.}, ``Blockchain for internet of things: A survey,''
  \emph{IEEE Internet of Things Journal}, vol.~6, no.~5, pp. 8076--8094, Oct.
  2019.

\bibitem{Ref_BH_nguyen2020privacy}
T.~{Nguyen}, N.~{Tran}, L.~{Loven}, J.~{Partala}, M.~{Kechadi}, and
  S.~{Pirttikangas}, ``Privacy-aware blockchain innovation for 6g: Challenges
  and opportunities,'' in \emph{2020 2nd 6G Wireless Summit (6G SUMMIT)}, 2020,
  pp. 1--5.

\bibitem{Ref_MAH_Minerva2020DitialTwin}
R.~{Minerva}, G.~M. {Lee}, and N.~{Crespi}, ``Digital twin in the iot context:
  A survey on technical features, scenarios, and architectural models,''
  \emph{Proceedings of the IEEE}, vol. 108, no.~10, pp. 1785--1824, 2020.

\bibitem{Ref_MAH_Pires2019DigitalTwin}
F.~{Pires}, A.~{Cachada}, J.~{Barbosa}, A.~P. {Moreira}, and P.~{Leitão},
  ``Digital twin in industry 4.0: Technologies, applications and challenges,''
  in \emph{2019 IEEE 17th International Conference on Industrial Informatics
  (INDIN)}, vol.~1, 2019, pp. 721--726.

\bibitem{Ref_MAH_Barricelli2019DigitalTwin}
B.~R. {Barricelli}, E.~{Casiraghi}, and D.~{Fogli}, ``A survey on digital twin:
  Definitions, characteristics, applications, and design implications,''
  \emph{IEEE Access}, vol.~7, pp. 167\,653--167\,671, 2019.

\bibitem{Ref_BH_hu2020edge}
H.~{Hu} and C.~{Jiang}, ``Edge intelligence: Challenges and opportunities,'' in
  \emph{2020 International Conference on Computer, Information and
  Telecommunication Systems (CITS)}, 2020, pp. 1--5.

\bibitem{Ref_BH_han2020robustness}
B.~{Han}, S.~{Yuan}, Z.~{Jiang}, Y.~{Zhu}, and H.~D. {Schotten}, ``Robustness
  analysis of networked control systems with aging status,'' in \emph{IEEE
  INFOCOM 2020 - IEEE Conference on Computer Communications Workshops (INFOCOM
  WKSHPS)}, 2020, pp. 1360--1361.

\bibitem{Ref_BH_jiang2020ai}
Z.~{Jiang}, S.~{Fu}, S.~{Zhou}, Z.~{Niu}, S.~{Zhang}, and S.~{Xu},
  ``Ai-assisted low information latency wireless networking,'' \emph{IEEE
  Wireless Communications}, vol.~27, no.~1, pp. 108--115, 2020.

\end{thebibliography}

\end{document}